\documentstyle[12pt,epsfig]{article}
\newcommand{\Frac}[2]%
{{\textstyle \frac{\mbox{\footnotesize $#1$}\rule[-0.9mm]{0mm}{1mm}}%
{\mbox{\footnotesize $#2$}\rule{0mm}{3.1mm}}}}
\renewcommand{\thefootnote}{\fnsymbol{footnote}}

\newcommand{\footfrac}[2]%

\begin{document}
\begin{titlepage}
\vspace*{-12 mm}
\noindent
\begin{flushright}
\begin{tabular}{l@{}}
\end{tabular}
\end{flushright}
\vskip 12 mm
\begin{center}
{\large \bf
Towards an understanding of nucleon spin structure: \\
[3mm]
from hard to soft scales
}
\\[14 mm]
{\bf Steven D. Bass$^a$ \ and \ Christine A. Aidala$^b$}
\\[10mm]
{\em
$^a$
Institute for Theoretical Physics, \\
Universit\"at Innsbruck,
Technikerstrasse 25, Innsbruck, A 6020 Austria
\\[5mm]
}
\vspace{0.5cm}
{\em
$^b$
Department of Physics, University of Massachusetts Amherst, \\
Amherst, MA 01003, U.S.A.
\\[5mm]
}

\end{center}
\vskip 20 mm
\begin{abstract}
\noindent The workshop {\it The Helicity Structure of the Nucleon}
(BNL June 5, 2006) was organized as part of the 2006 RHIC \& AGS
Users' Meeting to review the status of the spin problem and future
directions. The presentations can be found at \cite{meeting}.
Recent data suggest small polarized glue and strangeness in the proton.
Here we present a personal summary of the main results and presentations.
What is new and exciting in the data, and what might this tell us about
the structure of the proton ?

\end{abstract}

\vspace{9.0cm}

\end{titlepage}
\renewcommand{\labelenumi}{(\alph{enumi})}
\renewcommand{\labelenumii}{(\roman{enumii})}
\renewcommand{\thefootnote}{\arabic{footnote}}

\newpage
\baselineskip=6truemm

\section{Introduction}

It is nearly 20 years since the European Muon Collaboration (EMC)
published their polarized deep inelastic measurement of the proton's
$g_1$ spin dependent structure function
and
the flavour-singlet axial-charge $g_A^{(0)}|_{\rm pDIS}$
\cite{emc}.
Their results suggested that the quarks' intrinsic spin
contributes little of the proton's spin.
The challenge to understand the spin structure of the proton
\cite{bassrmp,reya,rhicspin}
has inspired a vast programme of theoretical activity and new experiments at
CERN, DESY, JLab, RHIC and SLAC.
Where are we today ?

We start by recalling the $g_1$ spin sum-rules.

These are derived starting from the dispersion relation for polarized
photon-nucleon scattering and, for deep inelastic scattering,
the light-cone operator product expansion.
One finds that the first moment of  the $g_1$ structure function
is related
to the scale-invariant axial charges of the target nucleon by
\begin{eqnarray}
\int_0^1 dx \ g_1^p (x,Q^2)
&=&
\Biggl( {1 \over 12} g_A^{(3)} + {1 \over 36} g_A^{(8)} \Biggr)
\Bigl\{1 + \sum_{\ell\geq 1} c_{{\rm NS} \ell\,}
\alpha_s^{\ell}(Q)\Bigr\}
\nonumber \\
& &
+ {1 \over 9} g_A^{(0)}|_{\rm inv}
\Bigl\{1 + \sum_{\ell\geq 1} c_{{\rm S} \ell\,}
\alpha_s^{\ell}(Q)\Bigr\}  +  {\cal O}({1 \over Q^2})
 - \ \beta_1 (Q^2) {Q^2 \over 4 M^2}
.
\nonumber \\
\label{eqc50}
\end{eqnarray}
Here $g_A^{(3)}$, $g_A^{(8)}$ and $g_A^{(0)}|_{\rm inv}$ are the
isovector, SU(3) octet and scale-invariant  flavour-singlet axial
charges respectively. The flavour non-singlet $c_{{\rm NS} \ell}$
and singlet $c_{{\rm S} \ell}$ Wilson coefficients are calculable in
$\ell$-loop perturbative QCD \cite{Larin:1997}. The term $\beta_1
(Q^2) {Q^2 \over 4 M^2}$ represents a possible subtraction constant
from the circle at infinity when one closes the contour in the
complex plane in the dispersion relation \cite{bassrmp,fixedpole}.
If finite, the subtraction constant affects just the first moment
sum-rule.
For a leading-twist subtraction:
 $\beta_1 (Q^2) = O(1/Q^2)$
 as $Q^2 \rightarrow \infty$.
The first
moment of $g_1$ plus the subtraction constant, if finite, is equal
to the axial-charge contribution. The subtraction constant
corresponds to a real term in the spin-dependent part of the forward
Compton amplitude.

If one assumes no twist-two subtraction constant ($\beta_1 (Q^2) = O(1/Q^4)$)
then the axial charge contributions saturate the first moment
at leading twist.
The isovector axial-charge is measured independently in neutron
beta-decays
($g_A^{(3)} = 1.2695 \pm 0.0029$ \cite{PDG:2004})
and
the octet axial charge is extracted from hyperon beta-decays
($g_A^{(8)} = 0.58 \pm 0.03$ \cite{fec}).
From the first moment of $g_1$,
polarized deep inelastic scattering experiments have been interpreted
to imply a small value for the flavour-singlet axial-charge:
\begin{equation}
g_A^{(0)}\bigr|_{\rm pDIS} = 0.15 - 0.35
\label{eqa1}
\end{equation}
-- considerably less than the value of $g_A^{(8)}$. In the naive
parton model $g_A^{(0)}|_{\rm pDIS}$ is interpreted as the fraction
of the proton's spin which is carried by the intrinsic spin of its
quark and antiquark constituents. When combined with the octet axial
charge this value corresponds to a negative
strange-quark polarization
$\Delta s = {1 \over 3} (g_A^{(0)}|_{\rm pDIS} - g_A^{(8)})$:
\begin{equation}
\Delta s = -0.10 \pm 0.04
\label{eqa2}
\end{equation}
-- that is,
polarized in the opposite direction to the spin of the proton.
Relativistic quark models generally predict values $g_A^{(0)} \sim 0.6$
with little polarized strangeness in the nucleon \cite{bassrmp,Ellis:1974}.

The Bjorken sum-rule for the isovector part of $g_1$
\cite{Bjorken:1966}
\begin{equation}
\int_0^1 dx g_1^{p-n}
=
\frac{g_A^{(3)}}{6}
\left[1 - \frac{\alpha_s}{\pi} - 3.583 \left(\frac{\alpha_s}{\pi} \right)^2
        - 20.215 \left(\frac{\alpha_s}{\pi} \right)^3 \right]
\label{eqc58}
\end{equation}
has been confirmed in polarized deep inelastic scattering experiments at
the level of 10\%
\cite{bjsr}.

\section{The shape of $g_1$}

\begin{figure}
\centerline{\psfig{figure=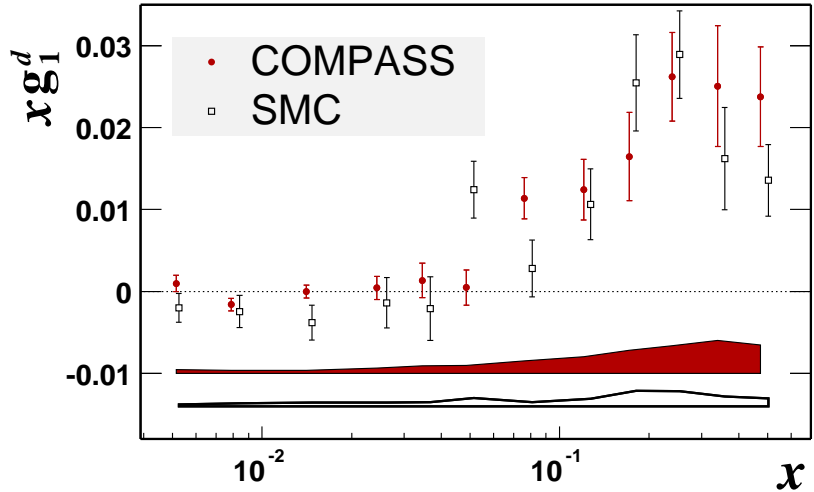,width=4.5in}}
\caption{Recent data on $g_1^d$ from COMPASS \cite{compass}.}
\label{fig:fig1}
\end{figure}

Deep inelastic measurements of $g_1$ have been performed in experiments at
CERN, DESY, JLab and SLAC.
There is a general consistency among all data sets.
COMPASS are yielding precise new data on $g_1^d$ at small $x$, down
to $x \sim 0.004$, shown in Fig.~\ref{fig:fig1} \cite{compass}. JLab
are focussed on the large $x$ region. To test deep inelastic
sum-rules it is necessary to have all data points at the same value
of $Q^2$. In the experiments the different data points are measured
at different values of $Q^2$, viz. $x_{\rm expt.}(Q^2)$.
Next-to-leading order (NLO) QCD-motivated fits taking into account
the scaling violations associated with perturbative QCD are
frequently used to evolve all the data points to the same $Q^2$. In
a recent fit reported at this meeting COMPASS evolve the world $g_1$
data set to a common value $Q^2=3$ GeV$^2$; a preliminary value was
quoted \cite{kabuss}
\begin{equation}
g_A^{(0)}|_{\rm pDIS} = 0.25 \pm 0.02 ({\rm stat.}) \pm {\rm ?}
\end{equation}
where the additional error denoted ``?'' reflects systematics and
theoretical error in the set up of the QCD-motivated fit.
Even more precise data are becoming available and were reported at
this meeting from COMPASS for the deuteron spin structure function
$g_1^d$ \cite{kabuss}.
The data show the remarkable feature that $g_1^d$
is
consistent with
zero in the small $x$ region between 0.004 and 0.02.

\begin{figure}
\centerline{\psfig{figure=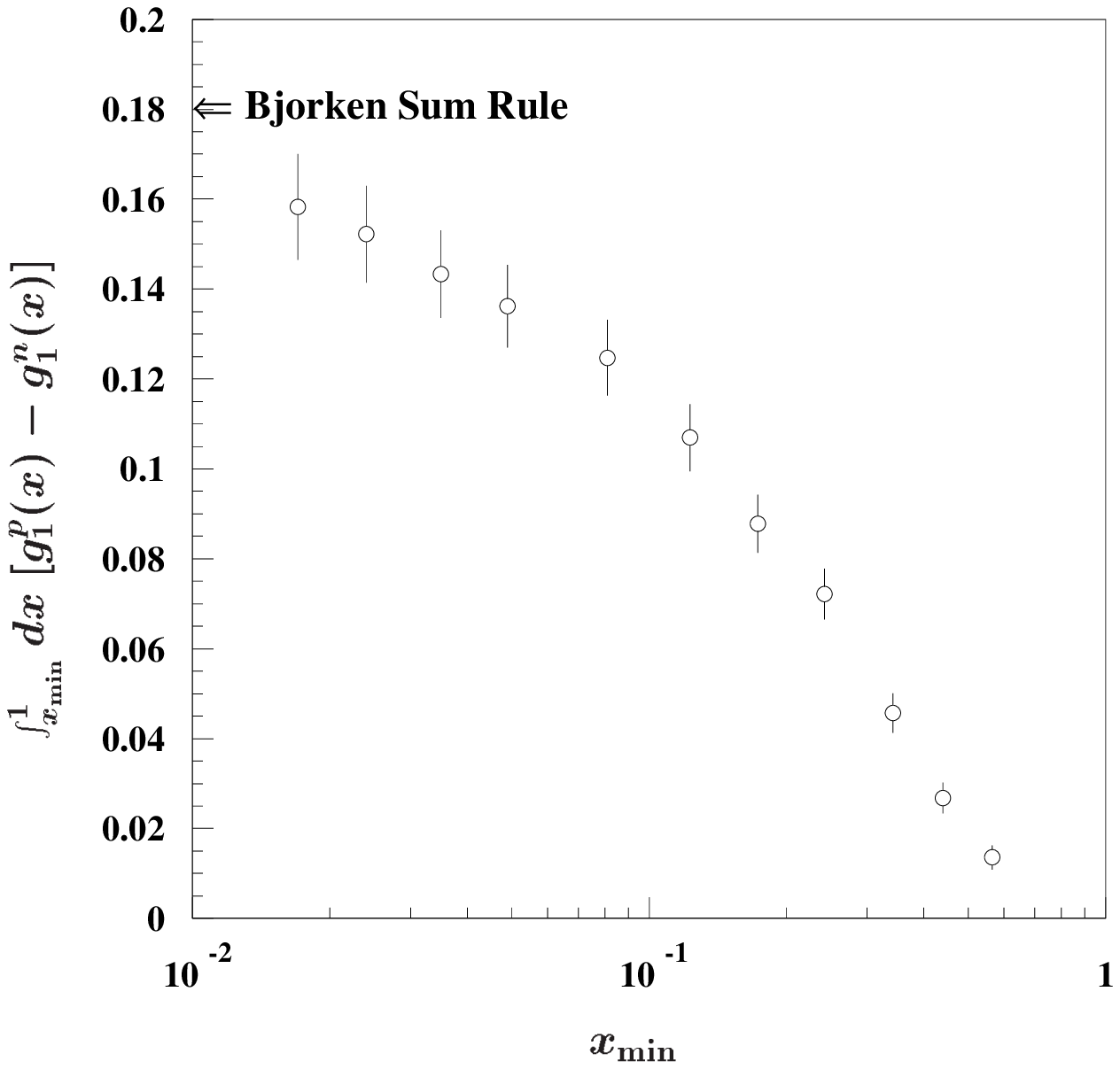,width=4.5in}}
\caption{Difference between the measured proton (SLAC E-143) and
neutron (SLAC E-154) integrals calculated from
a minimum $x$ value, $x_{\rm min}$ up to $x$ of 1 \cite{slac}.
The value is compared to
the theoretical prediction from the Bjorken sum rule which makes a prediction
over the full $x$ range.  For the prediction, the Bjorken sum rule is evaluated
up to third order in $\alpha_s$ and at $Q^2$ = 5 GeV$^2$.
}
\label{fig:fig2}
\end{figure}

In contrast, the isovector part of $g_1$ is observed to rise at
small $x$ ($0.01 < x < 0.1$) as $\sim x^{-0.5}$ and is much bigger
than the isoscalar part of $g_1$ \cite{bassrmp,bassmb}. This is in
sharp contrast to the situation in the unpolarized structure
function $F_2$ where the small $x$ region is dominated by isoscalar
pomeron exchange. The evolution of the Bjorken integral
$\int_{x_{\rm min}}^1 dx g_1^{p - n}$ as a function of $x_{min}$ is
shown for the SLAC data (E143 and E154) in Figure \ref{fig:fig2}
\cite{slac}. About 50\% of the sum-rule comes from $x$ values below
about 0.12 and about 10-20\% comes from values of $x$ less than
about 0.01 \cite{bassrmp}. The $g_1^{p-n}$ data are consistent with
quark model and perturbative QCD predictions in the valence region
$x > 0.2$ \cite{epja}. The size of $g_A^{(3)}$ forces us to accept a
large contribution from small $x$ and the observed rise in $g_1^{p -
n}$ is required to fulfil this non-perturbative constraint.

It would be interesting to extend precision measurements of the isovector
$g_1^{p-n}$ to smaller values of $x$ to further test its
low $x$ behaviour and to observe the convergence of the Bjorken integral
as a function of $x_{\rm min}$.
Possible measurements could be made at COMPASS running on a proton target
to complement the precise new deuteron-target data or with a future polarized
$ep$ collider.

The rise in $g_1^{p-n}$ is a challenge for Regge predictions and
perturbative QCD.
The Regge prediction for $g_1^{p-n}$ at small $x$ is
\begin{equation}
g_1^{p-n} \sim \sum_i f_i \ \frac{1}{x}^{\alpha_i}
.
\end{equation}
Here the $\alpha_i$ denote the Regge intercepts
for
isovector $a_1$ Regge exchange
and the $a_1$-pomeron cuts \cite{Heimann:1973}.
The coefficients $f_i$ are to be determined from experiment.
Soft Regge predictions for the leading $a_1$ intercept
$\alpha_{a_1}$ lie between -0.4 and -0.2 \cite{bass06} within the
phenomenological range quoted in \cite{Ellis:1988}. For the value
-0.2 the effective intercept corresponding to the $a_1$ soft-pomeron
cut is $\simeq -0.1$.
The $a_1$ and $a_1$ soft-pomeron cut alone are unable to account
for the $g_1^{p-n}$ data.
Does the rise in $g_1^{p-n}$ follow from $a_1$ exchange plus
perturbative QCD evolution {\it or} is there a distinct hard
exchange \cite{bass06} ?
-- that is, a polarized analogue of the one {\it or} two
pomerons question in unpolarized deep inelastic scattering
\cite{twopomerons} !
One possibility is an $a_1$ hard-pomeron cut,
with intercept $\simeq +0.2$,
in conjunction with QCD Counting Rules
factors still at work in the measured $x$ range.

If Regge intercepts are $Q^2$-independent, as suggested by
analyticity in $Q^2$ \cite{Cudell:1999}, the hard exchange observed
in $g_1^{p-n}$ should also contribute in the transition region and
in polarized photoproduction as well as in the spin-dependent part
of the proton-proton total cross-section \cite{bass06}. The latter
could be investigated at RHIC using Roman Pot detectors, e.g. using
the pp2pp apparatus, with a spin rotator before the detector to
achieve longitudinal polarization and varying over the energy range
of the machine. One would be looking for a leading behaviour $\Delta
\sigma^{p-n} \sim s^{-0.5}$ to $\sim s^{-0.8}$ instead of the simple
$a_1$ prediction $\sim s^{-1.4}$. The strategy would be to look for
a finite asymmetry at the lowest energy and, if a signal is found,
to keep measuring with increasing energy until the asymmetry
disappears within the experimental uncertainties. These measurements
would provide a valuable test of spin-dependent Regge theory
\cite{Heimann:1973,reggespin}. The leading non-perturbative
gluon-exchange contribution in the isoscalar part of $\Delta \sigma$
is expected to behave as $\sim (\ln s/\mu^2) / s$ where $\mu \sim
0.5 - 1$~GeV is a typical hadronic scale \cite{reggespin}.
High-energy polarized photoproduction and the transition region
could be investigated using a polarized electron-proton collider
\cite{bassadr,trento} or perhaps through measurement of low $Q^2$
asymmetries at COMPASS using a proton target.
Knowledge of spin-dependent Regge behaviour would help to constrain
the high-energy part of the Gerasimov-Drell-Hearn
sum-rule as well
as the high-energy extrapolations of
$g_1^{p-n}$ at intermediate $Q^2$ that go into the JLab
programme to extract information about higher-twist matrix elements
in the nucleon.

\begin{figure}[t!]
\vspace*{8.0cm}
\includegraphics{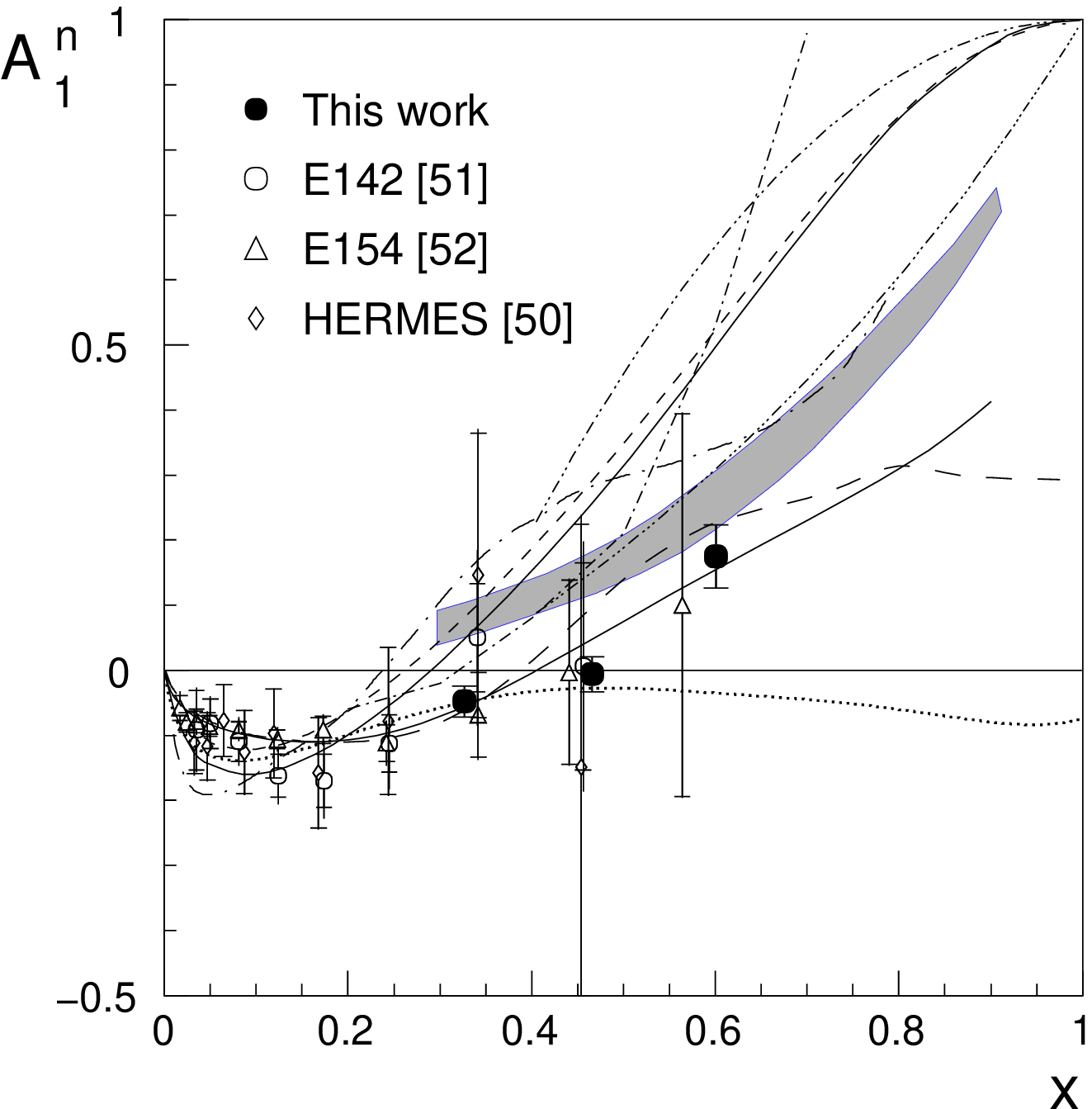}
\includegraphics{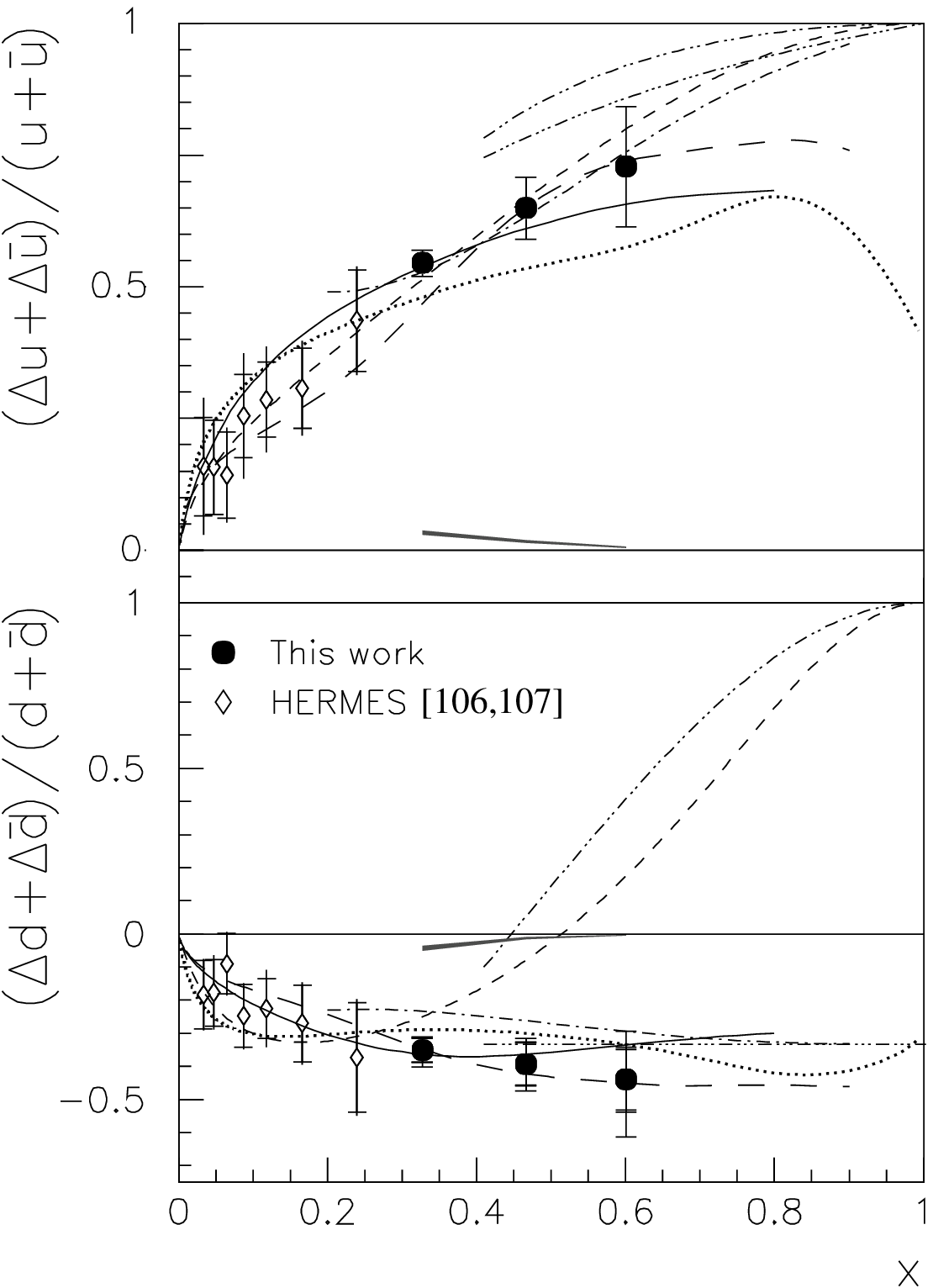}
\vspace*{4mm}
\caption[*]{Recent data on ${\cal A}_1^n$ from the Hall A E99-117
experiment at JLab (left) \cite{zheng}
and
extracted polarization asymmetries for $u+\bar{u}$
and $d+\bar{d}$ (right).
For more details and references on the various model predictions,
see \cite{zheng}.
\label{fig:fig3}}
\end{figure}

There are interesting puzzles also at large $x$. Recent data from
the Jefferson Laboratory Hall A Collaboration on the neutron
asymmetry ${\cal A}_1^n$ \cite{zheng} are shown in
Fig.~\ref{fig:fig3}. These data show a clear trend for ${\cal
A}_1^n$ to become positive at large $x$. The crossover point where
${\cal A}_1^n$ changes sign is particularly interesting because the
value of $x$ where this occurs in the neutron asymmetry is the
result of a competition between the SU(6) valence structure
\cite{Close:1988} and chiral corrections \cite{Schreiber:1988}.
Figure~\ref{fig:fig3} also shows the extracted flavour-dependent
asymmetries. The Hall A data are consistent with constituent quark
models with scalar diquark dominance which predict $\Delta d/d\to
-1/3$ at large $x$, while perturbative QCD Counting Rules
predictions (which neglect quark orbital angular momentum) give
$\Delta d/d\to 1$ and tend to deviate from the data, unless the
convergence to 1 sets in very late.

New preliminary Hall B data were reported at this meeting \cite{kuhn}
which appear to deviate from the Hall A measurements in
the larger $x$ region, $x \sim 0.6$, and are less inconsistent
with helicity Counting Rules predictions at large $x$.
We look forward to the final results and their extension to higher $x$.
A precision measurement of ${\cal A}_1^n$ up to $x \sim 0.8$ will
be possible following the 12 GeV upgrade of CEBAF and will provide
valuable input to resolving these issues.

Below scaling kinematics, JLab experiments are resolving the spin
structure of excited nucleon resonances and testing ideas about
quark-hadron duality \cite{kuhn}.

\section{Spin and the singlet axial charge}

There has been considerable theoretical effort to understand
the flavour-singlet axial-charge in QCD.
QCD theoretical analysis leads to the formula
\begin{equation}
g_A^{(0)}
=
\biggl(
\sum_q \Delta q - 3 {\alpha_s \over 2 \pi} \Delta g \biggr)_{\rm partons}
+ {\cal C}_{\infty}
.
\label{eqa10}
\end{equation}
Here $\Delta g_{\rm partons}$ is the amount of spin carried
by polarized
gluon partons in the polarized proton
($\alpha_s \Delta g \sim {\tt constant}$ as $Q^2 \rightarrow \infty$
\cite{etar})
and
$\Delta q_{\rm partons}$ measures the spin carried by quarks
and
antiquarks
carrying ``soft'' transverse momentum $k_t^2 \sim P^2, m^2$
where
$P$ is a typical gluon virtuality
and
$m$ is the light quark mass
\cite{etar,ccm}.
The polarized gluon term is associated with events in polarized
deep inelastic scattering where the hard photon strikes a
quark or antiquark generated from photon-gluon fusion and
carrying $k_t^2 \sim Q^2$ \cite{ccm}.
${\cal C}_{\infty}$ denotes a potential non-perturbative gluon
topological contribution
\cite{topology}
which is associated with the possible subtraction constant in
the
dispersion relation for $g_1$ \cite{bassrmp}.
If finite it would mean that
$\lim_{\epsilon \rightarrow 0} \int_{\epsilon}^1 dx g_1$
will measure
the difference of
the singlet axial-charge and the subtraction constant contribution;
that is, polarized deep inelastic scattering measures the combination
$g_A^{(0)}|_{\rm pDIS} = g_A^{(0)} - C_{\infty}$.

Possible explanations for the small value of $g_A^{(0)}|_{\rm pDIS}$
extracted
from the polarized deep inelastic experiments
include
screening from positive gluon polarization \cite{etar},
negative strangeness polarization in the nucleon \cite{bek},
a subtraction at infinity in the dispersion relation for $g_1$
\cite{bassrmp}
associated with non-perturbative gluon topology \cite{topology},
and
connections to axial U(1) dynamics \cite{tgv,bass99}.

There is presently a vigorous programme to disentangle the different
contributions. Key experiments involve semi-inclusive polarized deep
inelastic scattering (COMPASS and HERMES) and polarized
proton-proton collisions (PHENIX and STAR at RHIC).

One would like to understand the dynamics which suppresses the
singlet axial-charge extracted from polarized deep inelastic
scattering relative to the OZI prediction $g_A^{(0)} = g_A^{(8)}
\sim 0.6$ and also the sum-rule for the longitudinal spin structure
of the nucleon
\begin{equation}
{1 \over 2} = {1 \over 2} \sum_q \Delta q + \Delta g + L_q + L_g
\end{equation}
where $L_q$ and $L_g$ denote the orbital angular momentum contributions.
The theoretical basis for spin sum-rules for longitudinal and
transversely polarized targets is discussed in \cite{leader,leadersr}.

\begin{itemize}
\item
{\bf NLO QCD motivated fits to $g_1$}

The first attempts to extract information about gluon polarization
in the polarized nucleon used next-to-leading order (NLO)
QCD-motivated fits to inclusive $g_1$ data.

\begin{figure}[!t]
\vspace*{5.5cm} \includegraphics{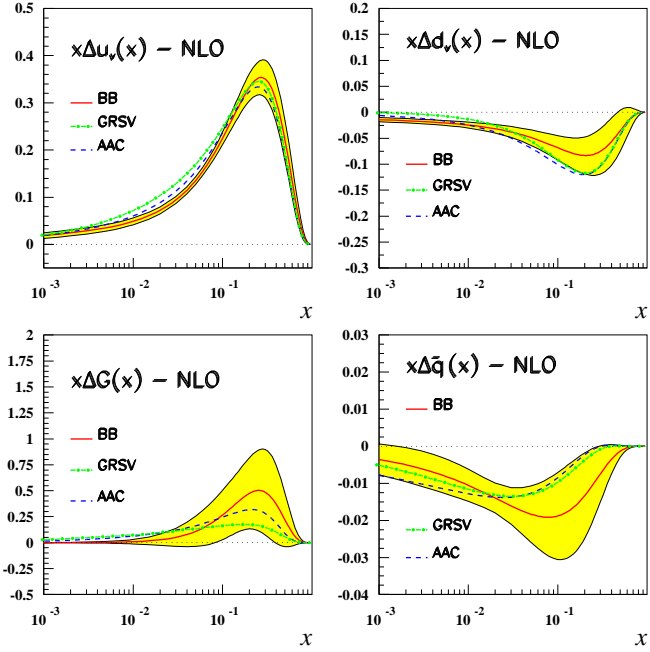} \vspace*{4.2cm}
\caption[*]{Polarized parton distribution functions from NLO pQCD
(${\overline {\tt MS}}$) fits at $Q^2 =4~$GeV$^2$ using SU(3)
flavour assumptions \cite{Stoesslein:2002}. \label{fig:fig4}}
\end{figure}

Similar to the analysis that is carried out on unpolarized data, global
NLO perturbative QCD analyses
have been performed on the polarized structure function data sets.
The aim is to extract the polarized quark and gluon parton distributions.
These QCD fits are performed within a given factorization scheme,
e.g.
the ``AB'', chiral invariant (CI) or JET and $\overline{\rm MS}$ schemes.
New fits are now being produced taking into account all
the available
data including new data from polarized semi-inclusive
deep inelastic scattering.
Typical polarized distributions extracted from the fits
are shown in Fig.~\ref{fig:fig4}.
Given the uncertainties in the fits
associated in part with the ansatz
chosen for the shape of the spin-dependent quark and gluon
distributions at a given input scale,
values of $\Delta g$
are extracted ranging between about zero and +2.
A recent COMPASS fit to the world data using the $\overline{\rm MS}$
scheme was
reported at this meeting.
Preliminary values extracted
for the polarized quark and gluon spin contributions are \cite{kabuss}
\begin{equation}
\Delta \Sigma = 0.25 \pm 0.02 ({\rm stat.}) \pm ?,
\ \ \ \ \ \
\Delta g = 0.4 \pm 0.2 ({\rm stat.}) \pm ?
\end{equation}
respectively,
where the additional error denoted
``?'' again reflects systematics
and theoretical error in the set up of the QCD motivated fit.

To go further more direct measurements involving glue sensitive
observables are needed to really extract the magnitude of
$\Delta g$ and
the shape of $\Delta g (x, Q^2)$
including any possible nodes in the distribution function.

\item
{\bf Gluon polarization}

There is a vigorous and ambitious global programme to measure
$\Delta g$. Interesting channels include gluon mediated processes in
semi-inclusive polarized deep inelastic scattering (COMPASS) and
hard QCD processes in high energy polarized proton-proton collisions
at RHIC.

\begin{figure}[t!]
\includegraphics{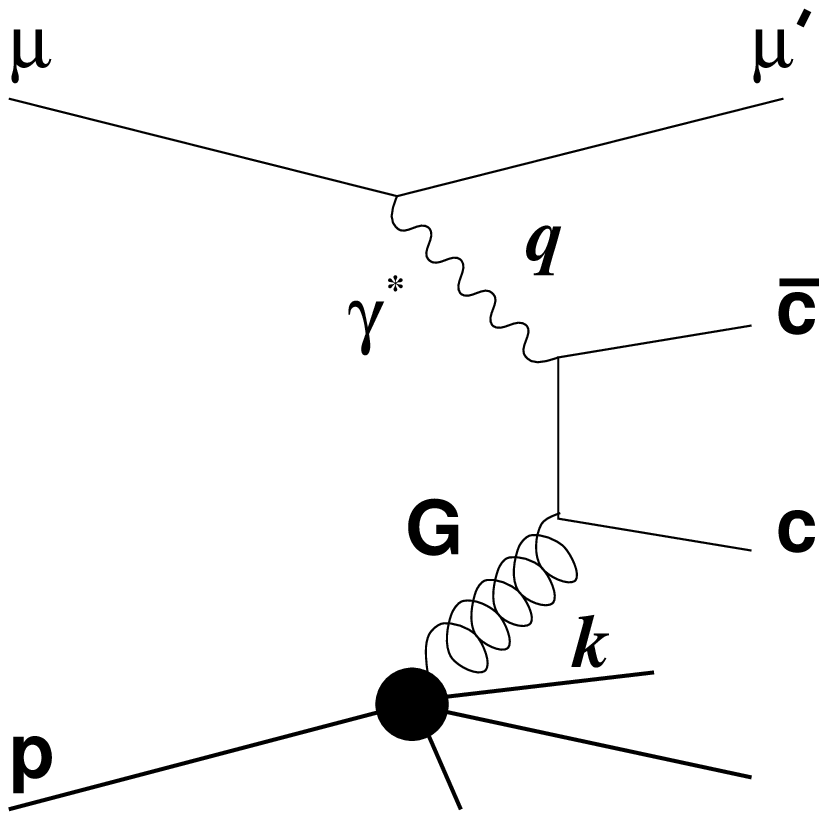}
\vspace{8.5cm}
\parbox{16.0cm}
{\caption[Delta]
{c $\overline{\rm c}$ production in Photon Gluon Fusion}
\label{fig:fig5}}
\end{figure}

COMPASS has been conceived to measure $\Delta g$ via the study of
the photon-gluon fusion process, as shown in Fig.~5. The cross
section for this process is directly related to the gluon density at
the Born level. The experimental technique consists of the
reconstruction of charmed mesons in the final state. COMPASS also
use the same process with high $p_t$ particles instead of charm to
access $\Delta g$ \cite{compassdeltag}. This leads to samples with
larger statistics but these have larger background contributions
from QCD Compton processes and fragmentation. High $p_t$ charged
particle production has been used in earlier attempts by HERMES
\cite{hermesdeltag} and SMC \cite{smcdeltag} to access gluon
polarization. These measurements together with preliminary results
reported at this meeting \cite {kabuss} are listed in Table 1 for
$x_g \sim 0.1$. An improvement of a factor of 2 in statistics is
anticipated from the 2006 COMPASS run in most channels.

\begin{table}[b!]
\caption{\label{tab:table1}
Polarized gluon measurements from deep inelastic experiments
}
\vspace{3ex}
\begin{tabular}{lccr}
  Experiment  &  process            &  $\langle x_g \rangle$
&  $\Delta g / g$   \\
\hline

HERMES      &  high $p_t$ hadrons   & 0.17 &
$ 0.41 \pm 0.18 \pm 0.03$
\\
SMC         &  high $p_t$ hadrons   & 0.07 &  $ -0.20 \pm 0.28 \pm 0.10$
\\
COMPASS     & high $p_t$ hadrons, $Q^2 < 1$
&  $ 0.085 $
&  $0.016 \pm 0.058 \pm 0.55$ \\
COMPASS & high $p_t$ hadrons, $Q^2 > 1$  & $0.13 $ &
$0.06 \pm 0.31 \pm 0.06$ \ {\it (prelim.)} \\
COMPASS & charm & $0.15 $ & $-0.57 \pm 0.41 ({\rm stat.})$ \ {\it (prelim.)} \\
\end{tabular}
\end{table}

RHIC Spin \cite{rhicspin} is achieving polarized proton-proton
collisions at 200 GeV centre of mass energy and $\sim 60\%$
polarization.  There was additionally a brief run at 62.4 GeV this
year, during which the PHENIX experiment elected to take data with
longitudinally polarized collisions.  A first test run at 500 GeV
centre of mass is taking place in June 2006.

\begin{figure}[t!]
\includegraphics{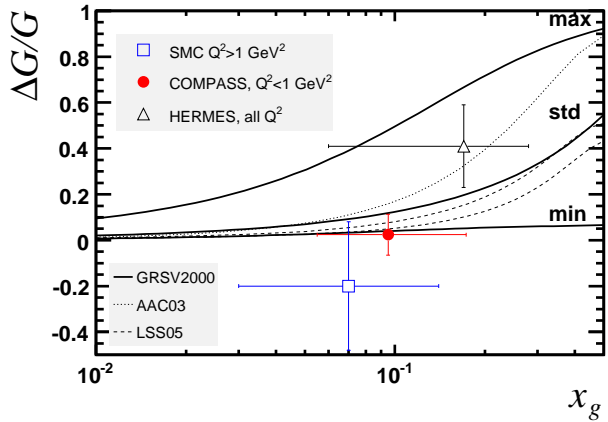}
\vspace{8.5cm}
\parbox{16.0cm}
{\caption[Delta]
{Polarized gluon measurements from COMPASS, HERMES and SMC
\cite{compassdeltag}. }
\label{fig:fig6}}
\end{figure}

\begin{figure}[t!]
\includegraphics{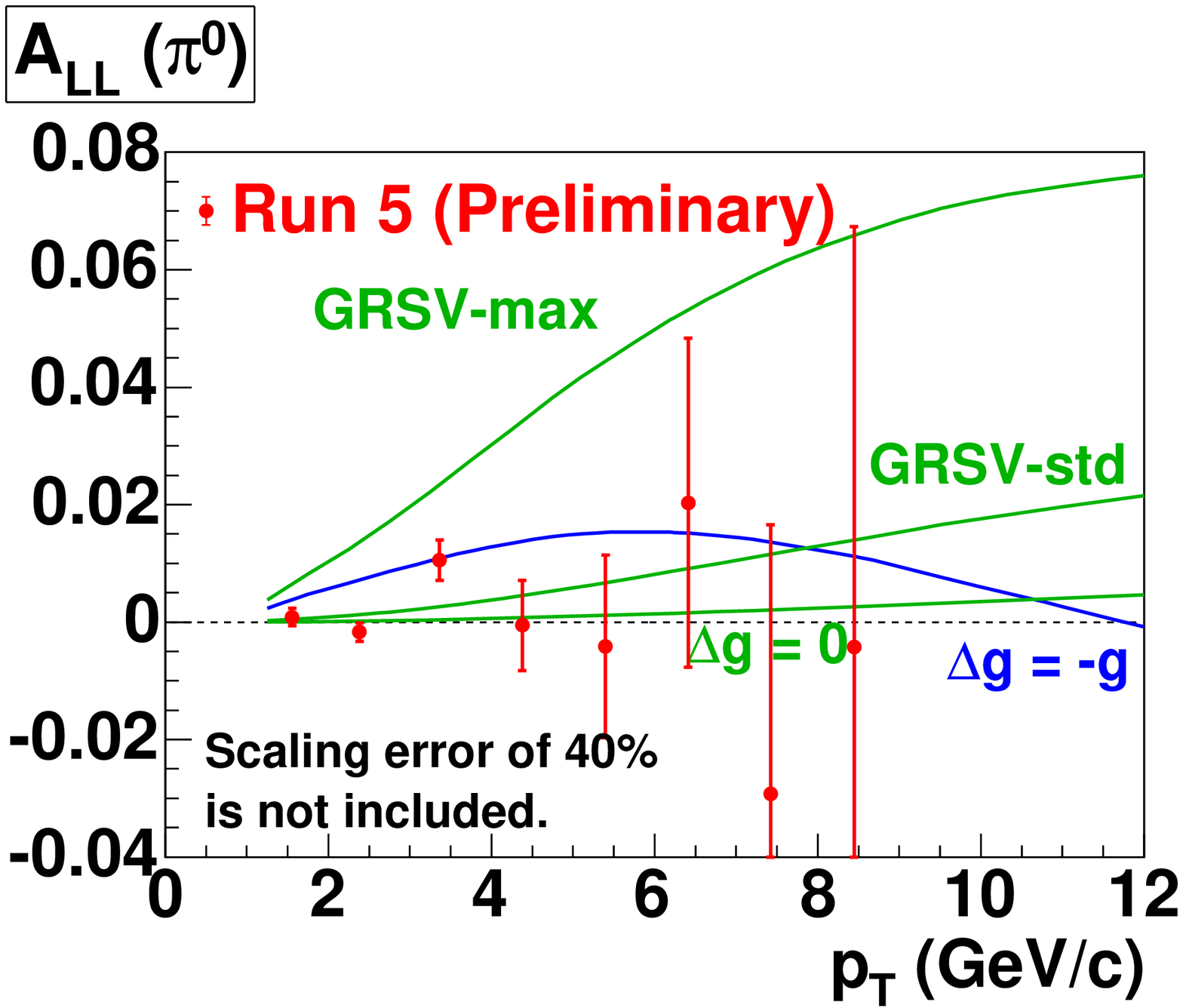}
\includegraphics{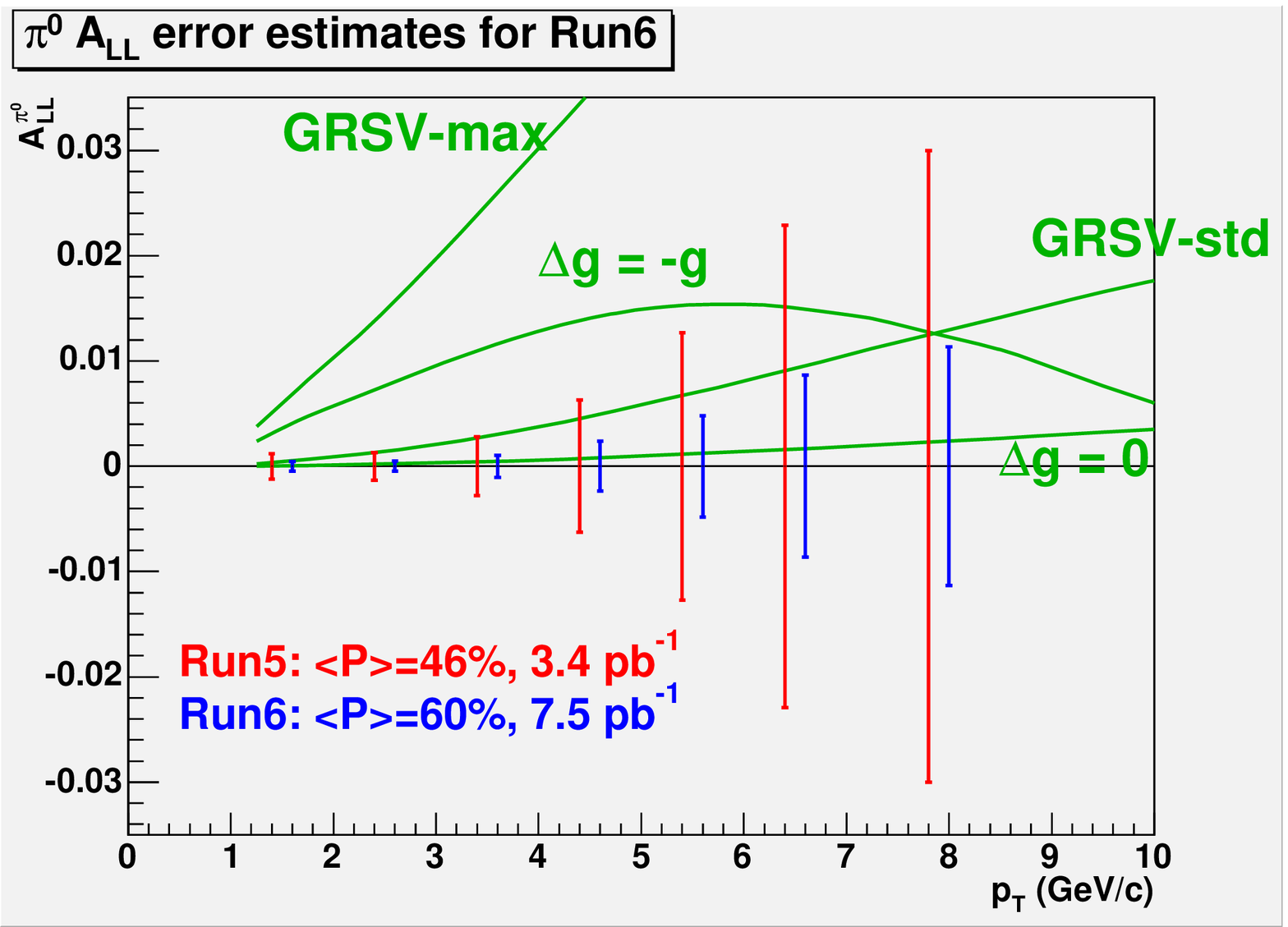}
\vspace{9.0cm}
\parbox{16.0cm}
{\caption[Delta]
{Preliminary PHENIX results on $A_{LL}^{\pi^0}$ together
 with the predictions from various QCD fits and (right)
 projections
 for the improvement in accuracy following the 2006 run
 \cite{younus,phenixdata}.}
\label{fig:fig7}}
\end{figure}
\begin{figure}[b!]
\includegraphics{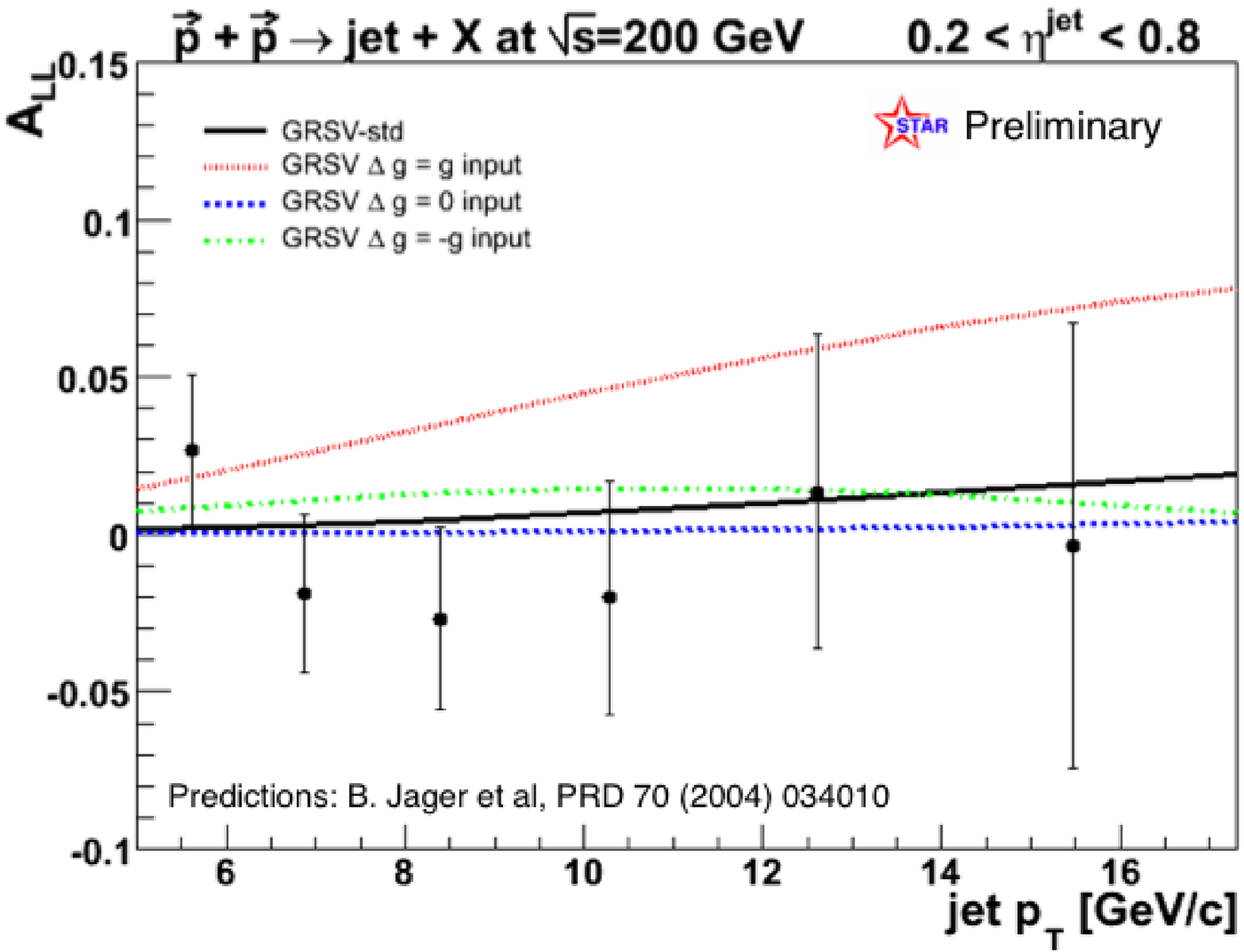}
\includegraphics{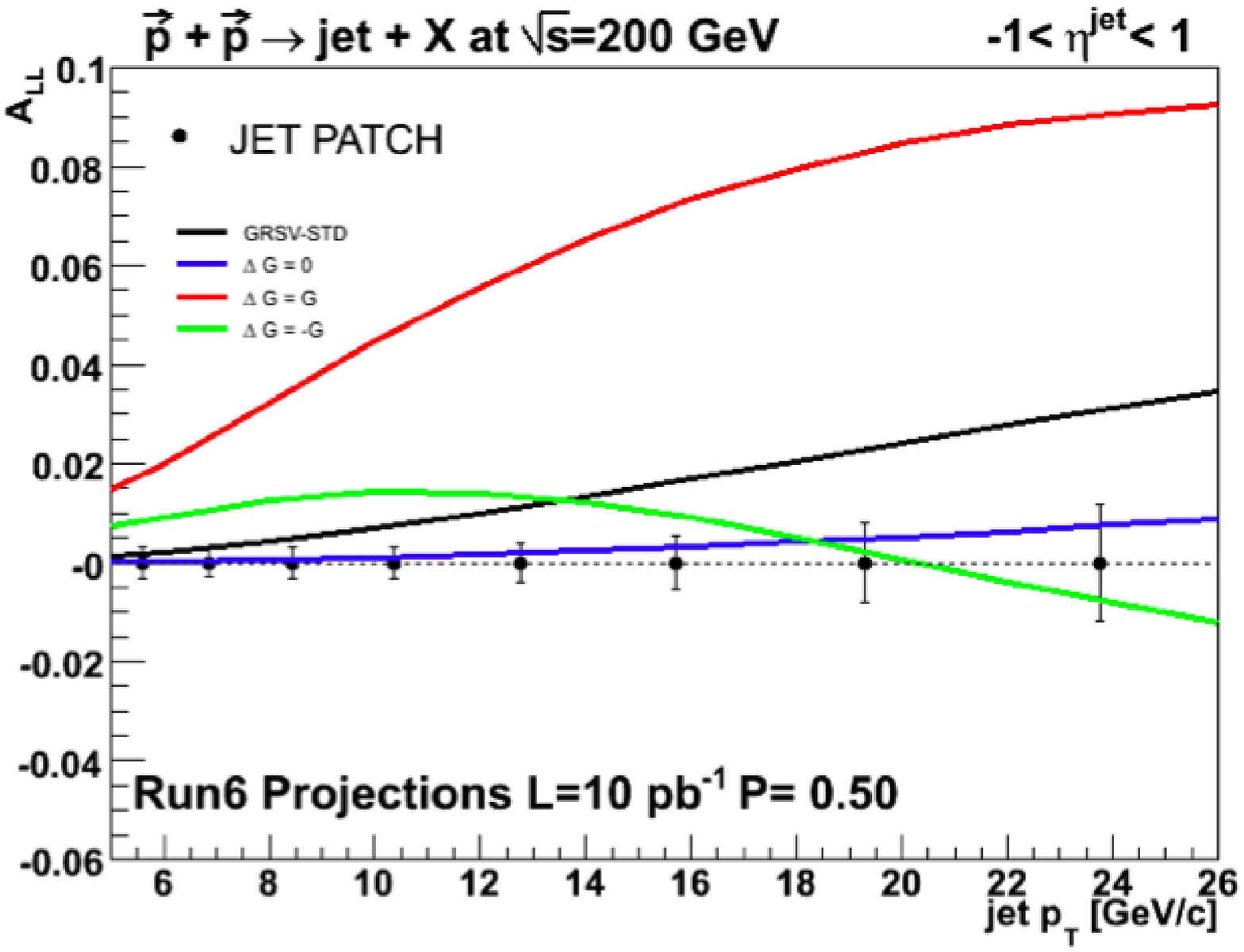}
\vspace{9.0cm}
\parbox{16.0cm}
{\caption[Delta]
{Preliminary STAR data on the longitudinal double spin inclusive jet
 asymmetry $A_{LL}$ for the years 2003-04 and (right) projections
 for the improvement in accuracy following the 2006 run
 \cite{sichtermann,stardata}.}
\label{fig:fig8}}
\end{figure}

The available data thus far into the RHIC spin programme and the
current PHENIX detector configuration have made the double-helicity
asymmetry of neutral pions the best probe of the gluon polarization
in PHENIX \cite{younus,phenixdata1,phenixdata}.
Charged pion asymmetries will
complement current measurements, with a significant measurement
expected by 2007.  Direct photons provide a theoretically cleaner
probe of $\Delta g$ and are directly sensitive to its sign but
require higher luminosity running.  The direct photon cross section
has already been measured \cite{phenixphotons}, with the first
asymmetry measurement expected from the 2005 data and a definitive
measurement at 200 GeV anticipated by 2009.  Future detector
upgrades will allow access to other probes sensitive to the gluon
polarization, such as open charm and jets.  In particular, a silicon
vertex barrel detector is planned for 2009, and a forward
calorimeter $(1 < |\eta| < 3)$ is planned for 2011.

An important channel at STAR providing sensitivity to the gluon is
jet production.  The 2003-2004 STAR measurement of the
double-helicity asymmetry in jet production
\cite{sichtermann,stardata} is expected to be greatly improved by
data from 2005 and 2006.  Charged pion asymmetries will provide
complementary sensitivity to gluon polarization; first results are
expected from 2005 data and will be further improved with 2006 data.
The mid-rapidity cross section for neutral pions at STAR was
recently released and is in good agreement with NLO pQCD
calculations.  This represents an important stepping stone for
future neutral pion and direct photon asymmetry measurements at
STAR, probing $\Delta g$.  Photon-jet correlations will provide
information on the kinematics of the partonic scattering.

The published data from COMPASS \cite{compassdeltag}, HERMES
\cite{hermesdeltag} and SMC \cite{smcdeltag} and the preliminary
data from PHENIX (05 run) \cite{younus,phenixdata} and STAR (03-04
runs) \cite{sichtermann,stardata} shown in this meeting appear in
Figs. 6, 7 and 8, together with the expectations of different NLO
fits to the inclusive $g_1$ data. Figures 7 and 8 also show
projections for the considerable improvement in accuracy expected in
the asymmetries following the successful 2006 run at RHIC
\cite{younus,sichtermann}. In Figs. 6-8 the curves ``GRSV-min'' (or
``$\Delta g=0$ input''), ``GRSV-std'', ``GRSV-max'' (or ``$\Delta
g=g$ input'') and ``$\Delta g = -g$ input'' \cite{grsv} correspond
to a first moment of
 $\Delta g \sim 0.1$, $0.4$, $1.9$ and $-1.8$ respectively at
$Q^2 \sim 1$ GeV$^2$
\cite{werner}.
Here ``input'' refers to
 the ``input scale'' $\mu^2 = 0.4$~GeV$^2$
 in the analysis of \cite{grsv}.

The COMPASS and RHIC Spin measurements suggest that polarized glue is,
by itself, not sufficient to resolve the difference between the small
value
of $g_A^{(0)}|_{\rm pDIS}$
and the constituent quark model prediction, $\sim 0.6$.
The
COMPASS data suggest that the gluon polarization is small
{\it or}
that it has a node in it around $x_g \sim 0.1$.
The PHENIX and STAR data are consistent with modest gluon polarization.
The considerable improvement in precision from the 2006 runs
at COMPASS and RHIC
should make it possible to resolve the different theoretical
expectations.
A combined NLO analysis of all the data would be valuable
\cite{forte} and, so far,
the COMPASS processes have been analysed only at leading order.
Nevertheless, the tentative conclusion
is that the gluon polarization may be small, $\ll 1$.
It is interesting to note that light-cone models \cite{bassbs,brodbs}
predict gluon
polarizations $\Delta g \sim 0.5 - 0.6$ at 1 GeV$^2$.
Further, the shape $\Delta g /g \sim x$ is expected on the basis of
only QCD Counting Rules at large $x$ plus colour coherence at small
$x$, with a value
$\Delta g/g \sim 0.105$ at $x_g \sim 0.1$ \cite{brodbs}
-- consistent with present data from COMPASS, HERMES and SMC.

\item
{\bf Sea polarization}

\begin{figure}[b!]
\vspace*{6.4cm} \includegraphics{hermes.epsi} \includegraphics{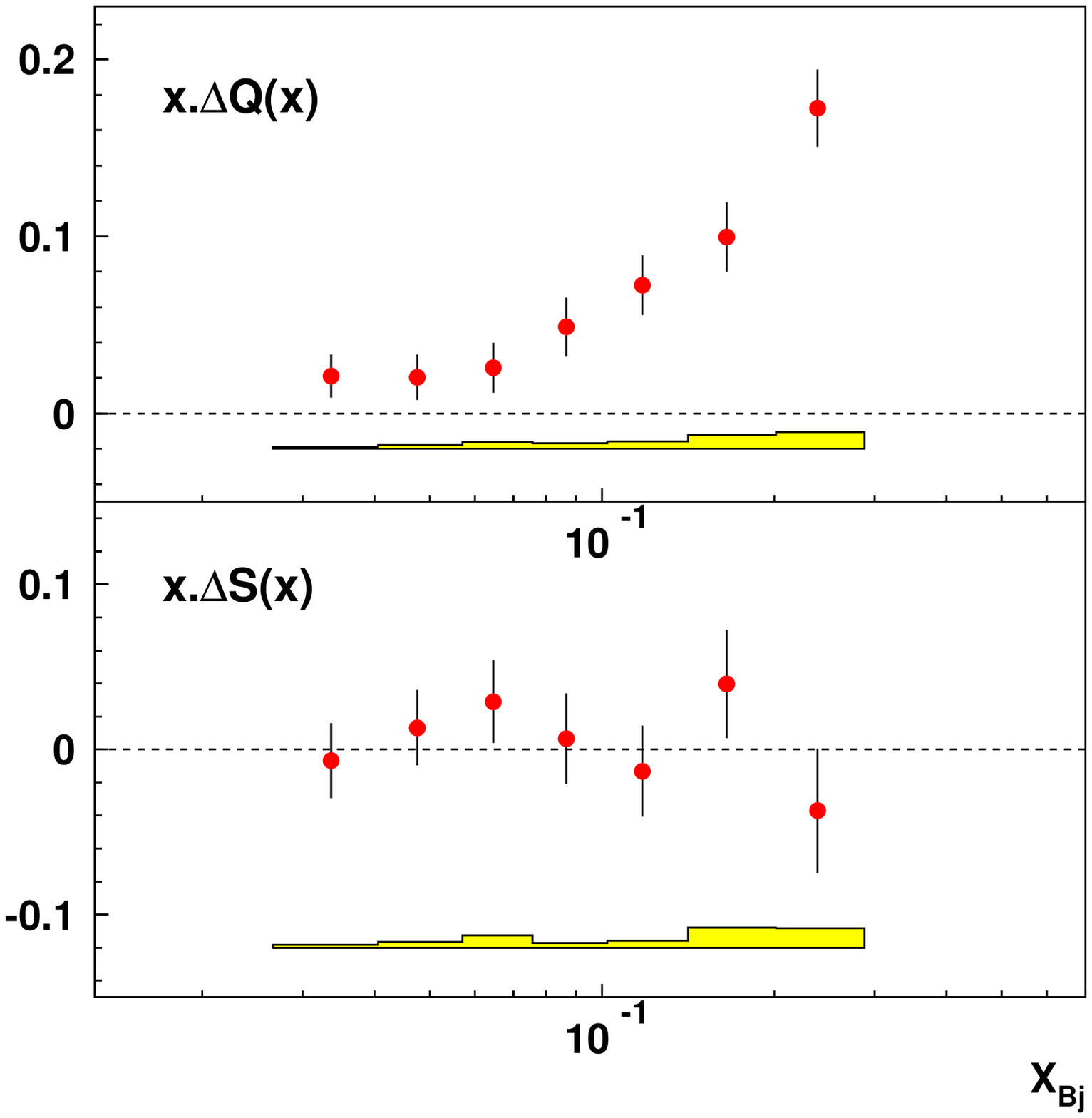}
\vspace*{3.7cm} \caption[*]{Recent {\sc Hermes} results for the
quark and antiquark polarizations extracted from semi-inclusive DIS.
Left: (a) the flavour separation reported in \cite{hermesdeltas};
Right: (b) the new preliminary results reported here \cite{liebing}
and in \cite{jackson}. Here $\Delta Q(x) = \Delta u(x) + \Delta
d(x)$. } \label{fig:fig9}
\end{figure}

Semi-inclusive measurements of fast pions and kaons in the current
fragmentation region with final state particle identification can be
used to reconstruct the individual up, down and strange quark
contributions to the proton's spin \cite{Close:1978}. In contrast to
inclusive polarized deep inelastic scattering where the $g_1$
structure function is deduced by detecting only the scattered
lepton, the detected particles in the semi-inclusive experiments are
high-energy (greater than 20\% of the energy of the incident photon)
charged pions and kaons in coincidence with the scattered lepton.
For large energy fraction $z=E_h/E_{\gamma} \rightarrow 1$ the most
probable occurrence is that the detected $\pi^{\pm}$ and $K^{\pm}$
contain the struck quark or antiquark in their valence Fock state.
They therefore act as a tag of the flavour of the struck quark
\cite{Close:1978}.

Figure~\ref{fig:fig9} shows the latest results on the flavour
separation from HERMES \cite{hermesdeltas}, which were obtained
using a leading-order Monte-Carlo code based ``purity'' analysis.
The polarizations of the up and down quarks are positive and
negative respectively, while the sea polarization data are consistent
with zero and not inconsistent with the negative sea polarization
suggested by inclusive deep inelastic data within the measured $x$
range \cite{grsv,Blumlein:2002}. However, there is also no evidence
from this semi-inclusive analysis for a large negative strange quark
polarization. For the region $0.023 < x < 0.3$ the extracted $\Delta
s$ integrates to the value $+0.03 \pm 0.03 \pm 0.01$ which contrasts
with the negative value for the polarized strangeness
(Eq.~\ref{eqa2}) extracted from inclusive measurements of $g_1$. In
a new analysis HERMES combine the inclusive deuteron asymmetry and
semi-inclusive kaon asymmetries to make a new extraction of $\Delta
s$. The analysis uses just isospin invariance and the charge
conjugation properties of the fragmentation functions. The
preliminary results \cite{liebing,jackson} are shown in Fig. 9b, and
the extracted $\Delta s$ is again consistent with zero.
It will be interesting to see whether this effect persists in
forthcoming semi-inclusive data from COMPASS.

For semi-inclusive hadron production experiments it is important to
match the theory with the acceptance of the detector
\cite{acceptance}. For example, the anomalous polarized gluon and
low $k_t$ sea contributions to $g_A^{(0)}$ in Eq.~(7) have different
transverse momentum dependence. The luminosity and angular
acceptance of the detector (150 mrad for HERMES) mean that these
semi-inclusive measurements of $\Delta s$ may be closer to $\Delta
s_{\rm partons}$ in Eq.~(7) than the inclusive value including the
polarized gluon term.

Spin transfer reactions also have the potential to provide insight into
QCD polarization phenomena.
Measurements of polarized (anti-)$\Lambda$ hyperon production are being
studied at
COMPASS, PHENIX and STAR as possible probes of strange quark polarization.

A direct and independent measurement of the strange quark axial-charge
through neutrino-proton elastic scattering
\cite{alberico,bcsta}, as proposed for JPARC and FNAL, would be valuable.
The axial charge measured in $\nu p$ elastic scattering is independent of
any assumptions about the presence or absence of a subtraction at
infinity in the dispersion relation for $g_1$ and the $x \sim 0$
behaviour of $g_1$.

The W programme at RHIC will provide flavour-separated measurements of
up and down quarks and antiquarks \cite{rhicspin}. A 500 GeV
commissioning run is planned for June 2006, and the high-energy
programme is expected to start in earnest in 2009.

Future neutrino factories would be an ideal tool for polarized
quark flavour decomposition studies.
These would allow one to collect large data samples of charged current
events, in the kinematic region $(x,Q^2)$ of present fixed target
data \cite{Forte:2001}. A complete separation of all four
flavours and anti-flavours would become possible, including $\Delta s(x,Q^2)$.

\end{itemize}

\section{Towards possible understanding }

Suppose that small gluon and strangeness polarization persist in future data.
Where will we be in our understanding of the (spin) structure of the proton
and the small value of $g_A^{(0)}|_{\rm pDIS}$ ?
The two possibilities would be {\it either}
a subtraction constant in the spin dispersion relation for $g_1$
{\it or} large SU(3) violation in the octet axial-charge
extracted from hyperon $\beta$-decays. The assumption of good SU(3)
is supported by the recent KTeV measurement \cite{KTeV} of the
$\Xi^0$ $\beta$-decay $\Xi^0 \rightarrow \Sigma^+ e^- {\bar \nu}$
and by recent theoretical analysis \cite{fec,ratcliffe}.
Further, a recent NLO analysis of inclusive and semi-inclusive
polarized deep inelastic data which allows
$g_A^{(8)}$ to float in a QCD-motivated fit reproduces the SU(3)
value $g_A^{(8)} = 0.58$ up to 8\% uncertainty \cite{deflorian}.

The total proton spin sum-rule for the sum of all quark and gluon
spin and orbital angular momentum contributions in Eq.~(8)
has to hold.
Relativistic motion which tends to shift some of the valence quark
total angular momentum from intrinsic spin to orbital angular
momentum acts equally in the iso-singlet
axial-charges $g_A^{(8)}$ and $g_A^{(0)}$ and cannot separate their values.

If there is a finite subtraction constant,
polarized high-energy processes are not measuring the full singlet
axial-charge:
$g_A^{(0)}$ and the partonic contribution $g_A^{(0)}|_{\rm pDIS}$
can be different.
Since the topological subtraction constant term affects just
the first
moment of $g_1$ and not the higher moments
it behaves like polarization at zero energy and zero momentum.

It is interesting to look for analogues in condensed matter physics.
Is there a system where the total spin is not just the sum of
the spin contributions of constituents carrying finite, non-zero,
momentum ?
Consider Helium-3 and Helium-4 atoms. These have the same chemical
structure and their properties at low temperatures are determined
just by their spins -- that is, the spin of the extra neutron in the
nucleus of the Helium-4 atom. The proton spin problem addresses the
question: Where does this spin come from at the quark level ?
In low temperature physics Helium-4 becomes a superfluid at 2K
whereas
Helium-3 remains as a normal liquid at these temperatures and
becomes superfluid only at 2.6 mK with a much richer phase diagram.
In the A-phase which forms at 21 bars pressure the spins of the
Cooper pairs align and a polarized condensate is formed.
The vacuum of the A-phase of superfluid Helium-3 behaves as an orbital
ferromagnet and uniaxial liquid crystal with spontaneous magnetisation
along the anisotropy axis ${\hat l}$
and as a spin antiferromagnet with
magnetic anisotropy along a second axis ${\hat d}$ \cite{anderson}.

In low energy processes the nucleon behaves like a colour-neutral
system of three massive constituent-quark quasi-particles
interacting self consistently with a cloud of virtual
Goldstone bosons (pions, ...)
and condensates
generated through dynamical chiral and axial U(1) symmetry breaking.
Suppose that the singlet component of the zero momentum ``condensate''
in the proton is spin polarized relative to the vacuum
outside the proton
with the polarization carried here by gluon topology \cite{topology}.
In this case the total singlet axial-charge,
as calculated in constituent quark models, would be the sum of the
partonic (finite momentum) and
``topological condensate'' (zero momentum) contributions.
The proton spin problem may be teaching us about
dynamical symmetry
breaking in QCD and the transition from current to constituent quarks.

\vspace{2.0cm}

{\bf Acknowledgements} \\

We thank the speakers, participants and organizers for a lively and
stimulating meeting.
We thank W. Guryn, E. Leader, H. Santos, E. Sichtermann and W. Vogelsang
for helpful conversations.
The research of
CAA is supported by the U.S. Department of Energy
(grant DE-FG02-88ER40415A022)
and
SDB is supported by the Austrian Science Fund (grant P17778-N08).
The meeting was supported by the U.S. Department of Energy Office of Science.

\end{document}